\documentclass[11pt,a4paper]{article}
\usepackage{epstopdf}
\usepackage{graphicx}
\usepackage{authblk}
\usepackage[hyperindex,breaklinks]{hyperref}

\newcommand{\beq}{\begin{equation}}
\newcommand{\eeq}{\end{equation}}
\newcommand{\beqa}{\begin{eqnarray}}
\newcommand{\eeqa}{\end{eqnarray}}
\newcommand{\beqan}{\begin{eqnarray*}} 
\newcommand{\eeqan}{\end{eqnarray*}}
\newcommand{\ba}{\begin{array}}
\newcommand{\ea}{\end{array}}

\title{Multigap RPC for PET: development and optimisation of the detector design}

\author{Georgi GEORGIEV$^{1}$, Nevena ILIEVA$^{2}$, Venelin KOZHUHAROV$^{1}$,  Iglika LESSIGIARSKA$^{1}$,  Leandar LITOV$^{1}$,  Borislav PAVLOV$^{1}$\thanks{Corresponding author: e-mail: Borislav.Pavlov@cern.ch}, Peicho PETKOV$^{1}$

\footnotesize
$^{1}$\textit{University of Sofia ``St. Kliment Ohridski'', Sofia, Bulgaria}\\
\footnotesize
$^{2}$\textit{Institute for Nuclear Research and Nuclear Energy, Sofia, Bulgaria}
}

\date{}

\begin{document}

\maketitle

\begin{abstract}Transforming the resistive plate chambers from charged-particle into gamma-quanta detectors opens the way towards their application as a basic element of a hybrid imaging system, which combines positron emission tomography (PET) with magnetic resonance imaging (MRI) in a single device and provides non- and minimally- invasive quantitative methods for diagnostics. To this end, we performed detailed investigations encompassing the whole chain from the annihilation of the positron in the body, through the conversion of the created photons into electrons and to the optimization of the electron yield in the gas. GEANT4 based simulations of the efficiency of the RPC photon detectors with different converter materials and geometry were conducted for optimization of the detector design. The results justify the selection of a sandwich-type gas-insulator-converter design, with Bi or Pb as converter materials.


\end{abstract}

\section{RPC as PET detectors --- a brief overview}
Positron Emission Tomography (PET) is a nuclear-medicine imaging technique for registration of whole-body distribution of positron-emitting biomarkers \cite{pet}. The emitted positrons annihilate and produce pairs of 511 keV photons, flying in opposite directions.  The PET registers these gamma quanta and reconstructs the so-called {\it line of response} (LOR). The standard PET devices use scintillating crystals  as photon detectors, coupled to photomultiplier tubes (PMT) or silicon photomultipliers (SiPM) in some advanced designs \cite{sipm, axpet}.

\begin{figure}[hb]
\begin{center}
\includegraphics[trim= 0 -0.4cm 0 -2.0cm, clip, width=.295\textwidth]{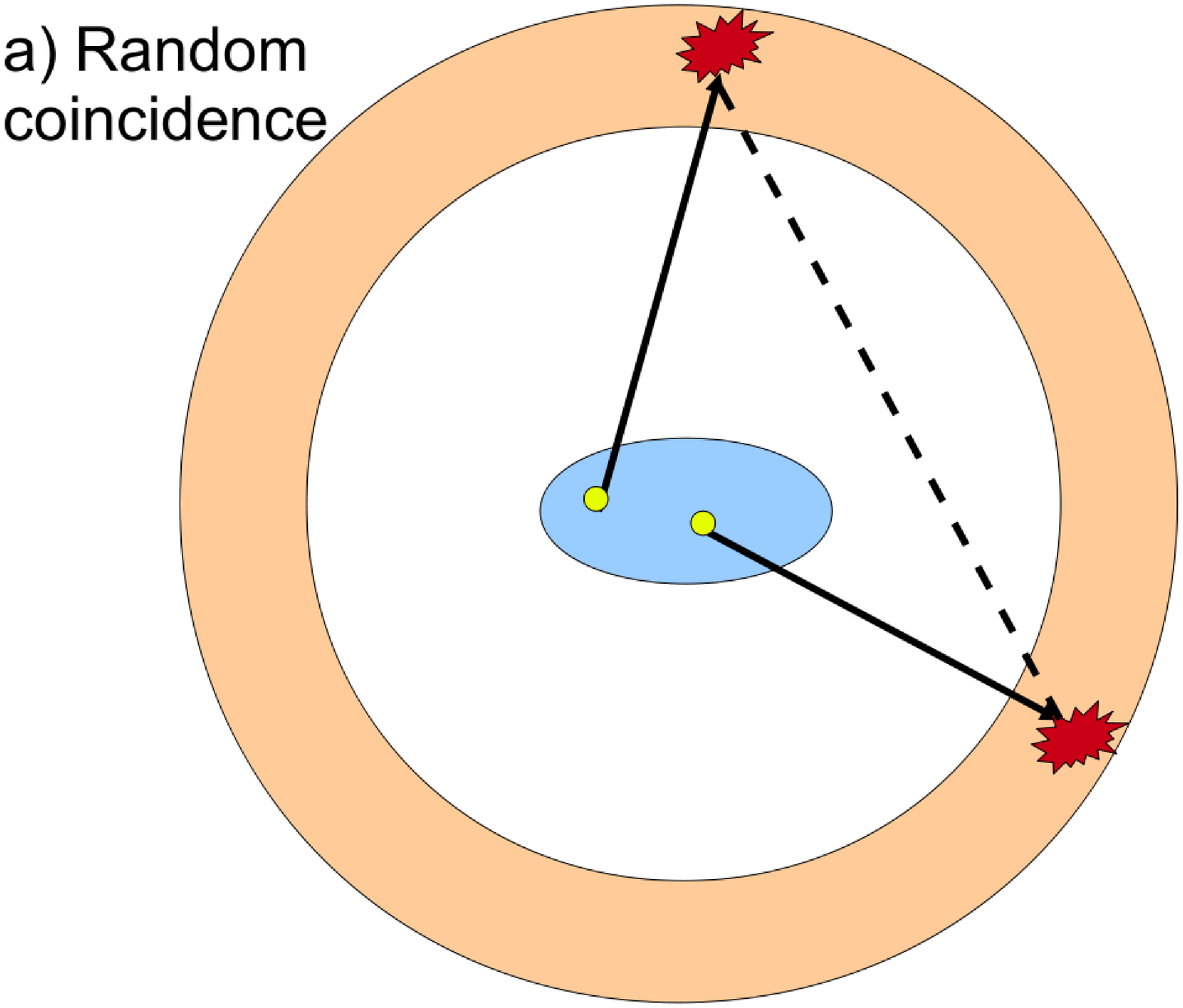}
\includegraphics[trim= 0 -0.4cm 0 -2.0cm, clip, width=.295\textwidth]{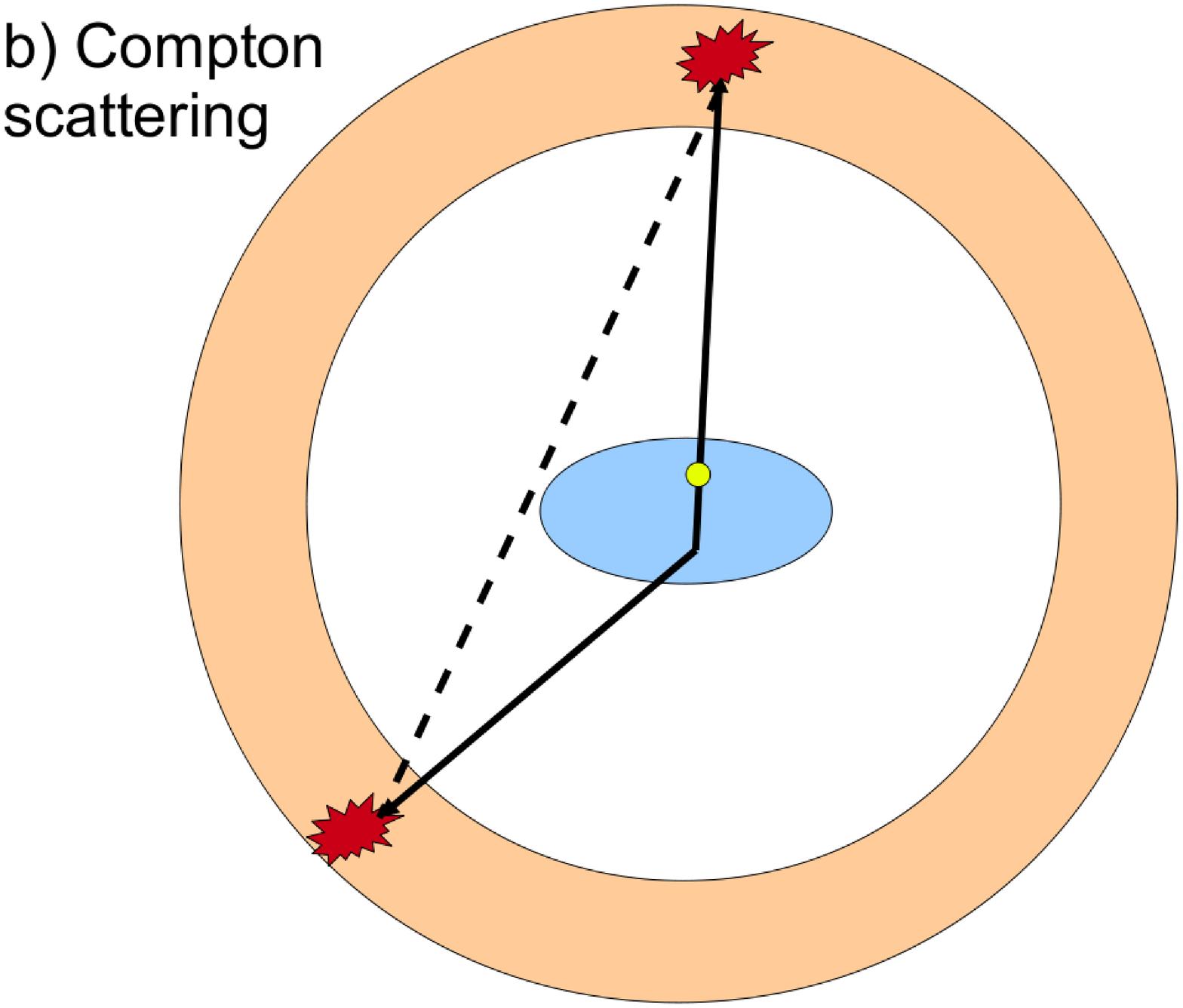}
\includegraphics[trim= 0 -0.4cm 0 -2.0cm, clip, width=.295\textwidth]{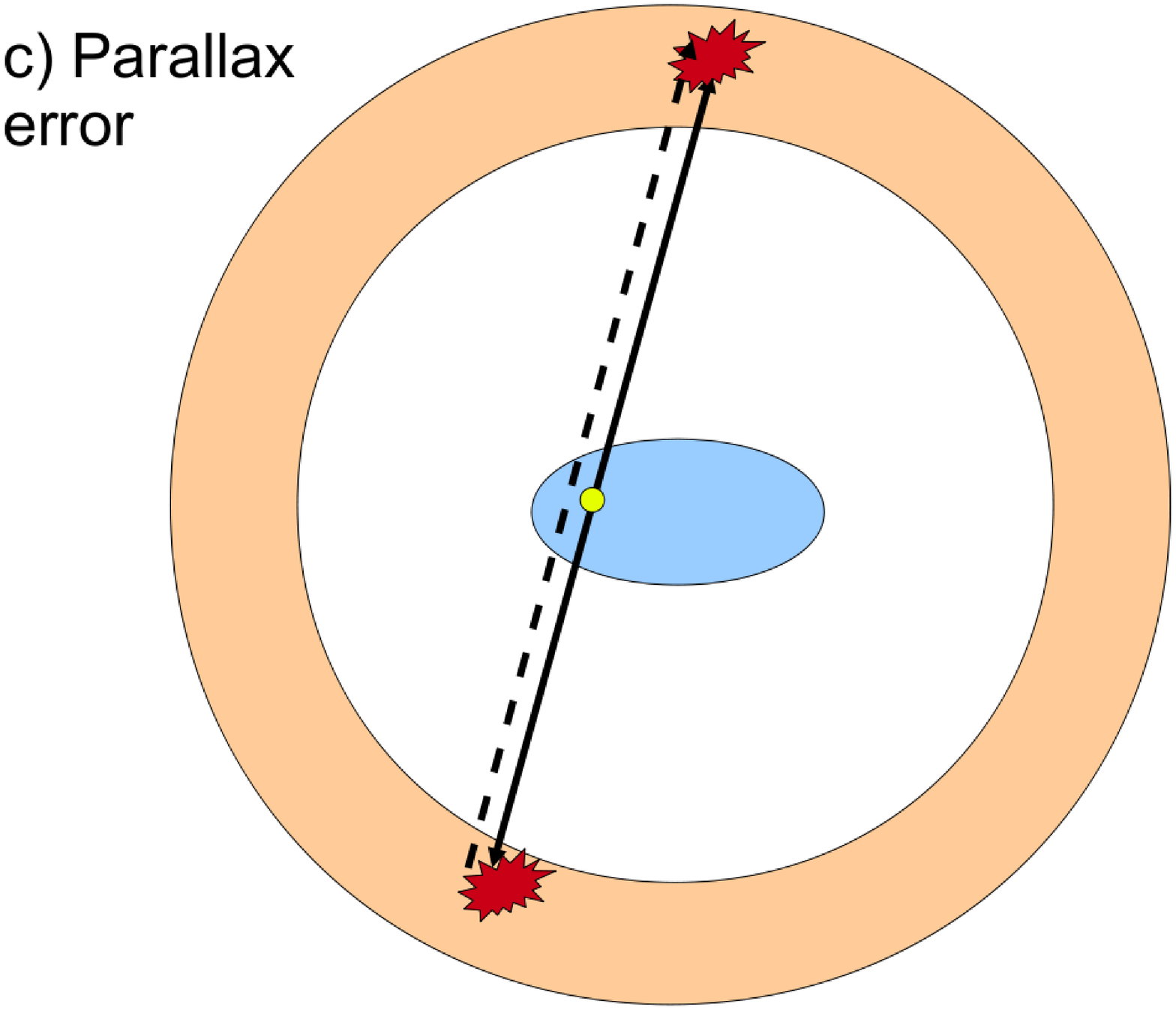}\\  
\hspace*{-0.2cm}(a)       
\hspace*{4.2cm}( b)      
\hspace*{4.0cm}( c)                                                                             
\end{center}
\caption{Processes that degrade PET resolution : a) random coincidence; b) Compton scattering; c) parallax error}
\label{fig1}
\end{figure}

\smallskip
The physical limitations of the PET image reconstruction accuracy are due to the registration of random coincidences and additionally scattered in the body photons, Fig.\ref{fig1} (a), (b). The former is proportional to the detector time window (the time period in which two registered photons are considered originating from the same annihilation event), thus it is determined by the detector time resolution. The latter is proportional to the detector sensitivity to photons with energies lower than 511 keV, as the additional scattering decreases the photon energy. Finally, the so called parallax error, Fig.\ref{fig1}(c), heavily depends on the detector spatial resolution as it accounts for the finite size of the individual detector elements (the detector ``pixel").

\smallskip
PET scan gives information about the density distribution and metabolism of the biomarker but does not provide a clear anatomical framing which is envisaged within hybrid imaging modalities, e.g. with computed tomography (PET/CT) or magnetic-resonance imaging  (PET/MRI)\cite{petctpetmri},
the latter being considered by many experts as the optimal diagnostic combination to become a real breakthrough in the clinical practice \cite{petmri}.  The main problem in the PET/MRI system is the sensitivity of the traditionally used photomultiplier tubes to magnetic fields. A prospective PET candidate for a hybrid PET/MRI system should be insensitive to strong and fast varying magnetic fields, as are  SiPM and dSiPM (digital silicon photomultipliers)\cite{dSiPM}.

\begin{figure}[ht]
\begin{center}
 \includegraphics[scale=0.8, width=10cm]{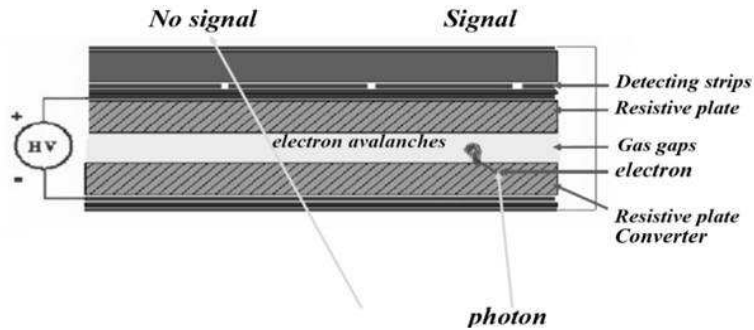}
\end{center}
\caption{Principal scheme of a PET-adapted RPC.}
\label{RPCPET}
\end{figure}

\smallskip
Resistive-plate chambers (RPC)\cite{santonico} are gaseous parallel-plate charged-particle detectors with plate resistivity of about 
$(10^{10}\div 10^{11})\Omega$ cm that are widely used in large-scale high energy physics experiments. Choosing appropriate materials for one or both of the electrodes transforms them into gamma-to-electron converters, Fig. \ref{RPCPET}, which in turn modifies RPC into a gamma-quanta detector. This idea was put forward already in \cite{fonte2}, thus making RPC an appealing alternative to the scintillating crystals \cite{rpcpet1, rpcpet2}. 

\smallskip
PET detectors that are based on resistive-plate chambers (RPC) do not encounter magnetic field compatibility problems and also minimize the image reconstruction inaccuracy discussed above. We aim at the development, justification and optimisation of a certain RPC-PET detector design and on the construction of a small-size and full-scale prototypes. 

\smallskip
The decisive RPC advantages from a PET point of view are:\vspace{-3pt}
\begin{itemize}\itemsep-2pt
\item excellent time resolution (20 ps time resolution reported for charged particles \cite{crispin});
\item sub-millimeter spatial resolution (see, e.g. \cite{rpcpet1});
\item absence of parallax error;
\item insensitivity to strong and varying magnetic fields, which ensures compatibility with MRI;
\item effective Compton-scattered photons suppression without energy measurement;
\item possibility for building detectors with a large field of view (FOV) \cite{cms, atlas};
\item substantially lower price in comparison to the crystals.

\end{itemize}
 
RPC's excellent time resolution opens the possibility to measure photons time of flight (TOF). The TOF information constrains the positron annihilation position to a few millimeters region on the LOR. It helps Compton suppression and also allows for reducing the acquisition time. Thus TOF measurement enhances the image reconstruction and essentially reduces the patient's dose\footnote{The typical PET dose is approximately 8 mSv, according to IAEA.}, opening the way to new PET medical applications in cardiology and neurology (see, e.g. \cite{fonte1}). The large field-of-view detectors enable also a simultaneous whole-body scan with a short acquisition time and an essentially lowered patient's dose.

\section{Prototype design and construction}

The photons from positron annihilation interact with the human body and with the converter medium of the PET detector through Compton scattering and photo-effect. In both cases the interaction cross-section is proportional to the atomic number $Z$ of the material (linearly, resp. to its fifth power) and anti-proportional to the energy of the photons. The ejected electrons eventually pass to the gas gap and develop an avalanche.

\medskip\noindent
{\bf Electron yield in the gas gap.}
The main objectives of the RPC-based PET design are maximum possible detector efficiency for 511 keV photons and suppression or rejection of Compton-scattered photons. The RPC efficiency is determined by the electron yield in the gas gap, given by the number of photons, for which at least one interaction within the converter has lead to the ejection of an electron into the gas gap. Electron yield depends on two processes: photon interactions in the converter and electron propagation through the converter to the gas. The electron distribution in the converter is given by 
\beq\label{e_distr}
\frac{dN}{dx}=kN_g-bN,
\eeq
where $x$ is the depth, $k$ is a photon interaction coefficient, $N_g$ is the number of photons at depth $x$, $b$ is an electron interaction coefficient.
 
For thin converters, when $N_g$ can be considered as a constant, the solution of Eq. (\ref{e_distr}) is:	
\beq\label{thin_conv}
N=a_b\left(1-e^{-x/b}\right),
\eeq
where $a_b$ is the maximum electron yield in the gas ($a_b = N_0 k/b$, $N_0$ is the initial number of photons), so the electron yield increases with $x$ till some maximum value, when saturation occurs.

In the case of larger converter thickness, photon-beam attenuation takes place and the solution of Eq. (\ref{e_distr}) reads instead:
\beq\label{thick_conv}
N=a_l\left(e^{-x/c}-e^{-x/b}\right),
\eeq
where $a_l$ is the maximum electron yield in the gas ($a_l = N_0 k/(b-1/c)$); $c$ is a coefficient which accounts for the photon beam attenuation. In this case, with the increasing of $x$ the electron yield decreases after the maximum.

Thus, on the one hand, the conversion probability increases with the converter thickness, but on the other hand, the electrons have a finite range in the converter medium. The optimal detector design requires a careful weighting of these two effects.
 
We investigated several different RPC-PET detector designs by means of GEANT4 \cite{geant4_1, geant4_2} simulations. The calculations were based on GEANT4 physical models for particle interactions at low energies. Because of the photon-interaction cross-sections increase with $Z$, elements with atomic numbers between 74 and 83 --- W, Pt, Au, Pb, and Bi --- were investigated as possible converters\footnote{The electron yield tends asymptotically to a maximum, Eq.(\ref{thin_conv}); in the analysis we refer to the converter thickness at which 95\% of the maximum value is reached.}. 

\begin{figure}[!h]
\begin{center}
\includegraphics[width=.495\textwidth]{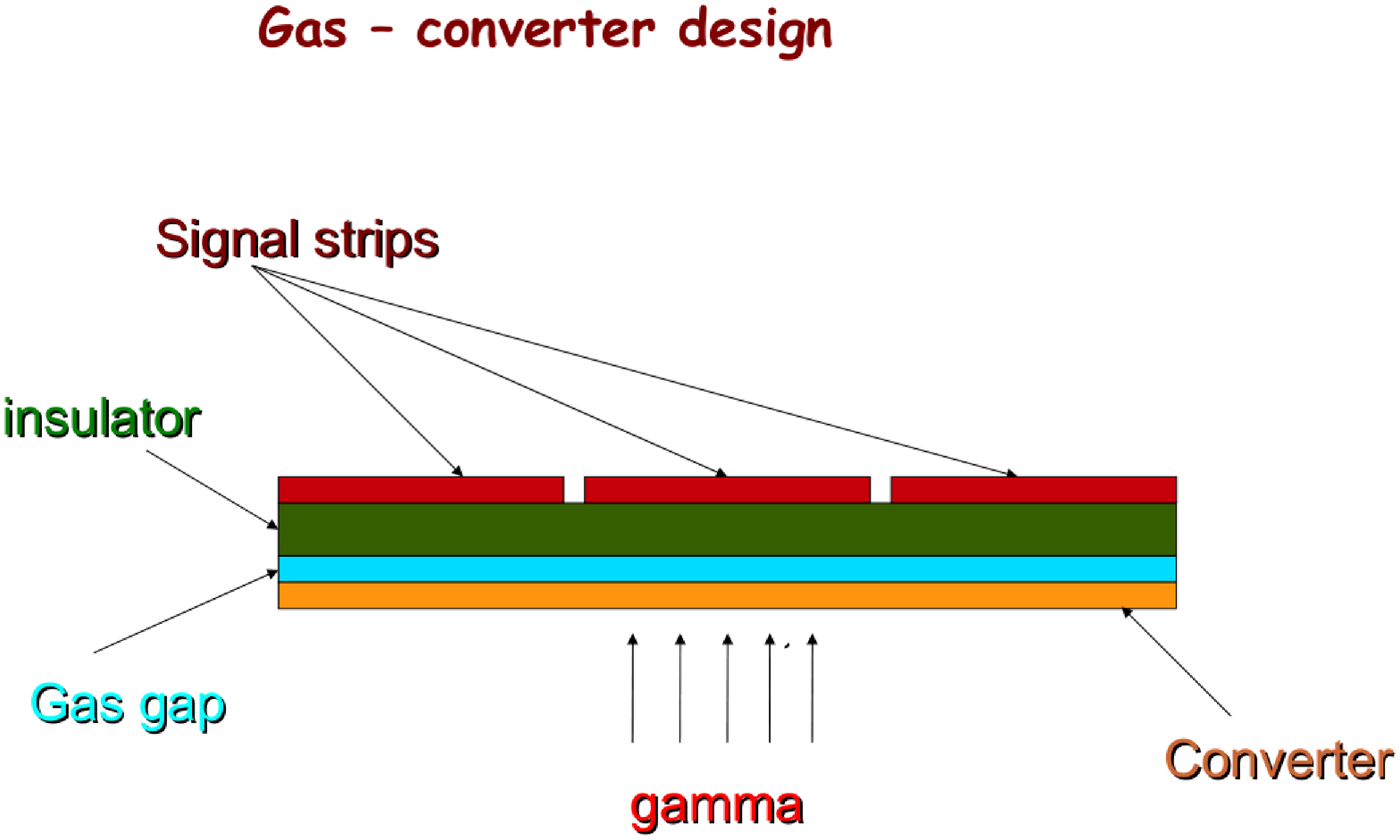}
\includegraphics[width=.495\textwidth]{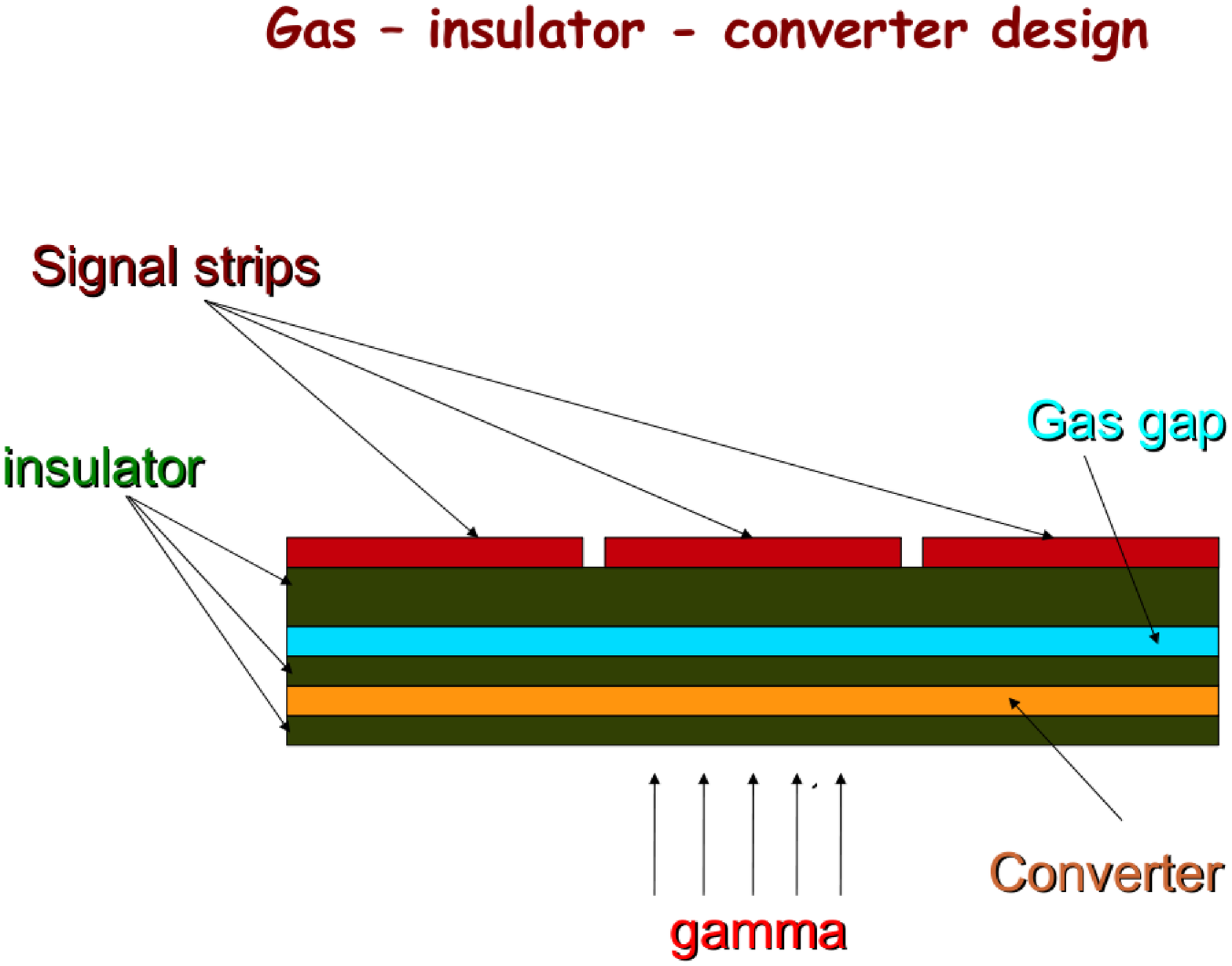}\\  \hspace*{0.1cm} (a) \hspace*{7cm} ( b)\\
\includegraphics[width=.495\textwidth]{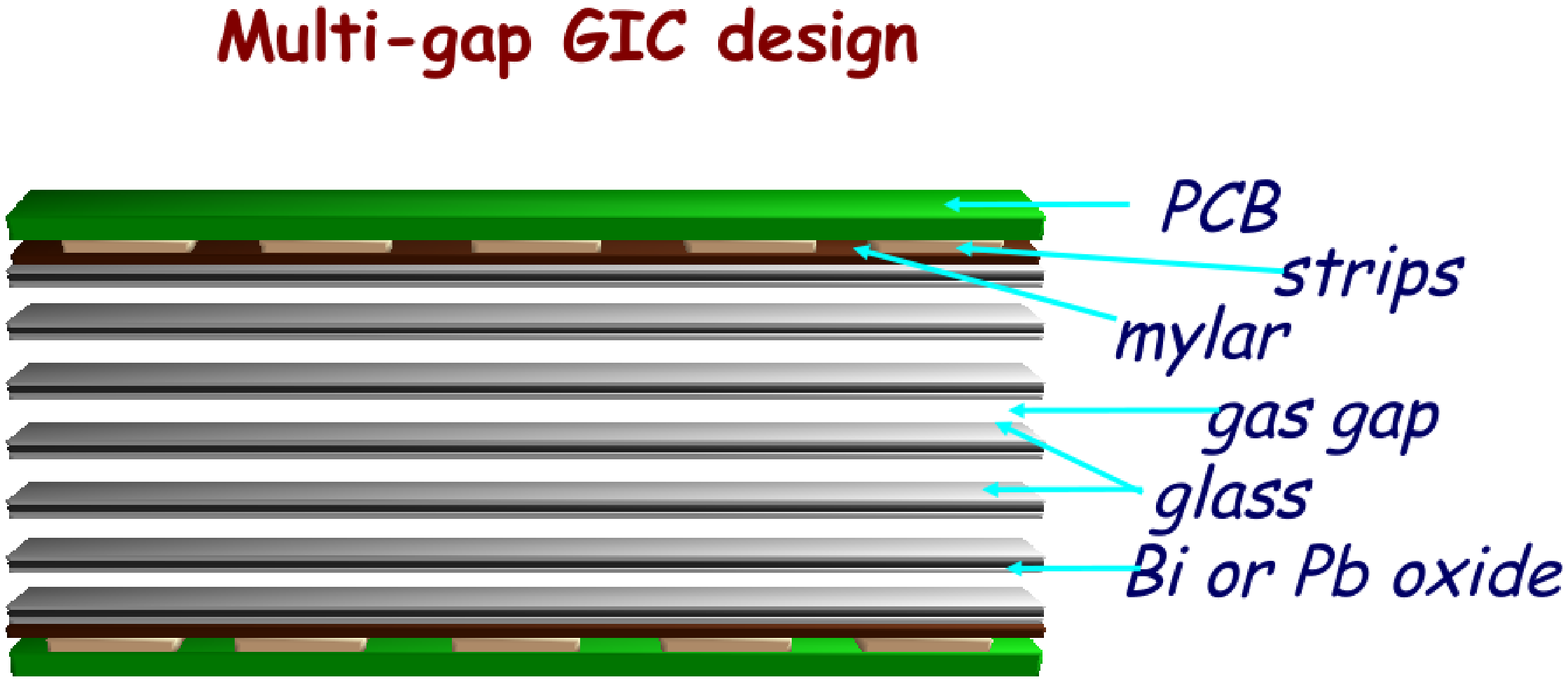}\\   \hspace*{-1.5cm} (c)
\end{center}
\caption{Possible RPC-PET designs: (a) gas-converter (GC) design; (b) gas-insulator-converter (GIC) design; (c) multigap gas-insulator-converter (MGIC) design.}
\label{fig2}
\end{figure}

\smallskip
\smallskip\noindent
{\bf Gas-converter (GC) design.} 
The direct contact between the converter and the gas apparently facilitates the propagation of the emitted electrons into the gas gap. The simulated GC design includes a 300 $\mu$m gas gap and 2 mm glass plate (Fig. \ref{fig2}a), with different converter materials and thicknesses. We studied five different high-$Z$ converting materials in a direct contact with the gas volume. One might expect problems due to the direct contact between the metal electrode and the gas, essentially metal aging and possible triggering of discharges in the RPC by excellent conductors like gold or platinum. This seems not to be the case, because only the resistivity of the anode is crucial  \cite{hppc}. Moreover, one can keep the converting properties of the material and greatly change the electrical properties. For example, lead has atomic number $Z=82$ and density $\rho = 11.34$ g cm$^{-3}$, lead oxide (PbO) has effective atomic number $Z_{eff} \approx 79.4 $ and density $\rho = 9.53$~g~cm$^{-3}$. Lead is a poor conductor, whereas lead oxide is an excellent insulator.

\begin{figure}[h!]
    \begin{center}
    \setlength{\unitlength}{1cm}   
    \includegraphics[height=5.5cm, width=6.5cm]{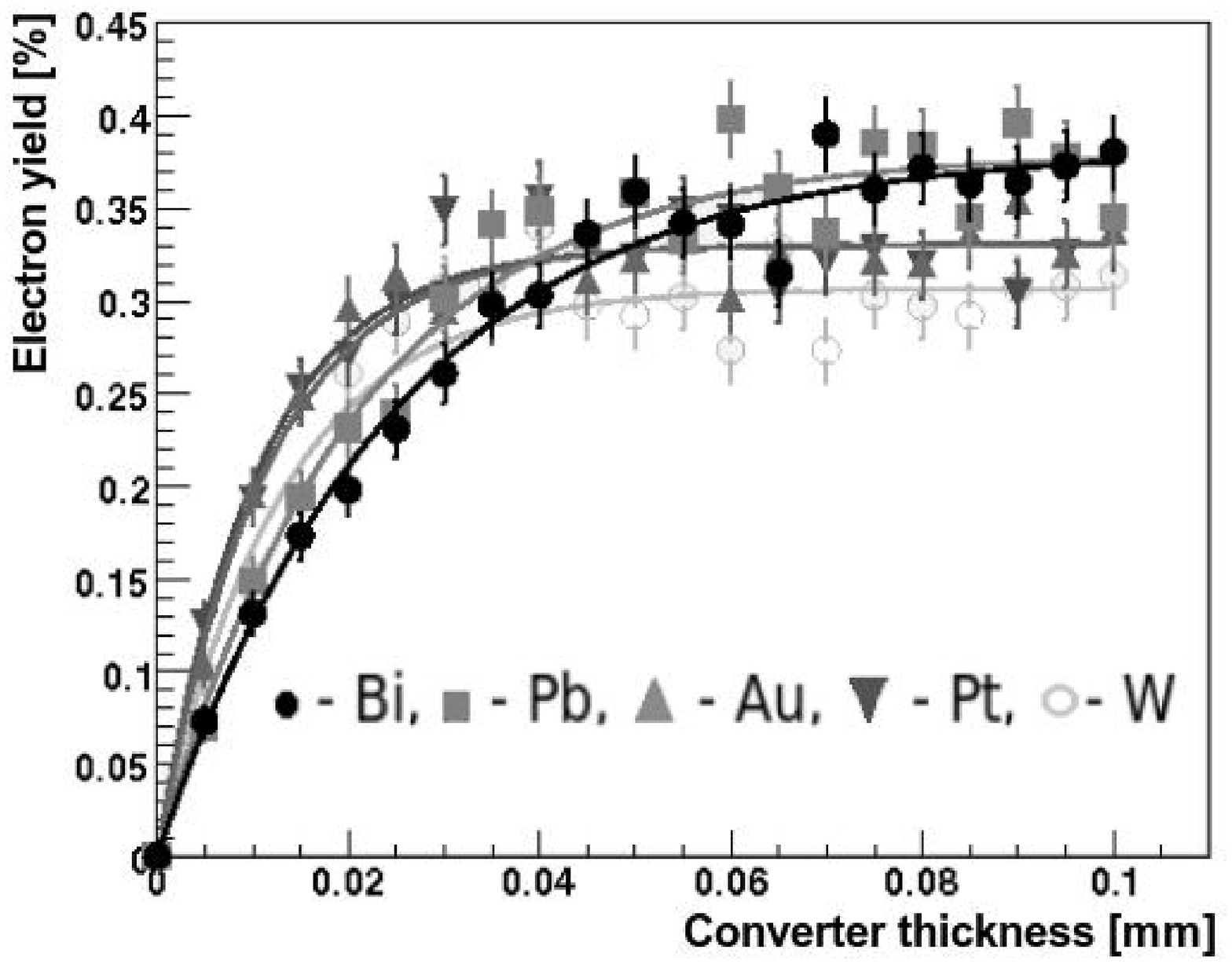}
    \includegraphics[height=5.5cm, width=6.5cm]{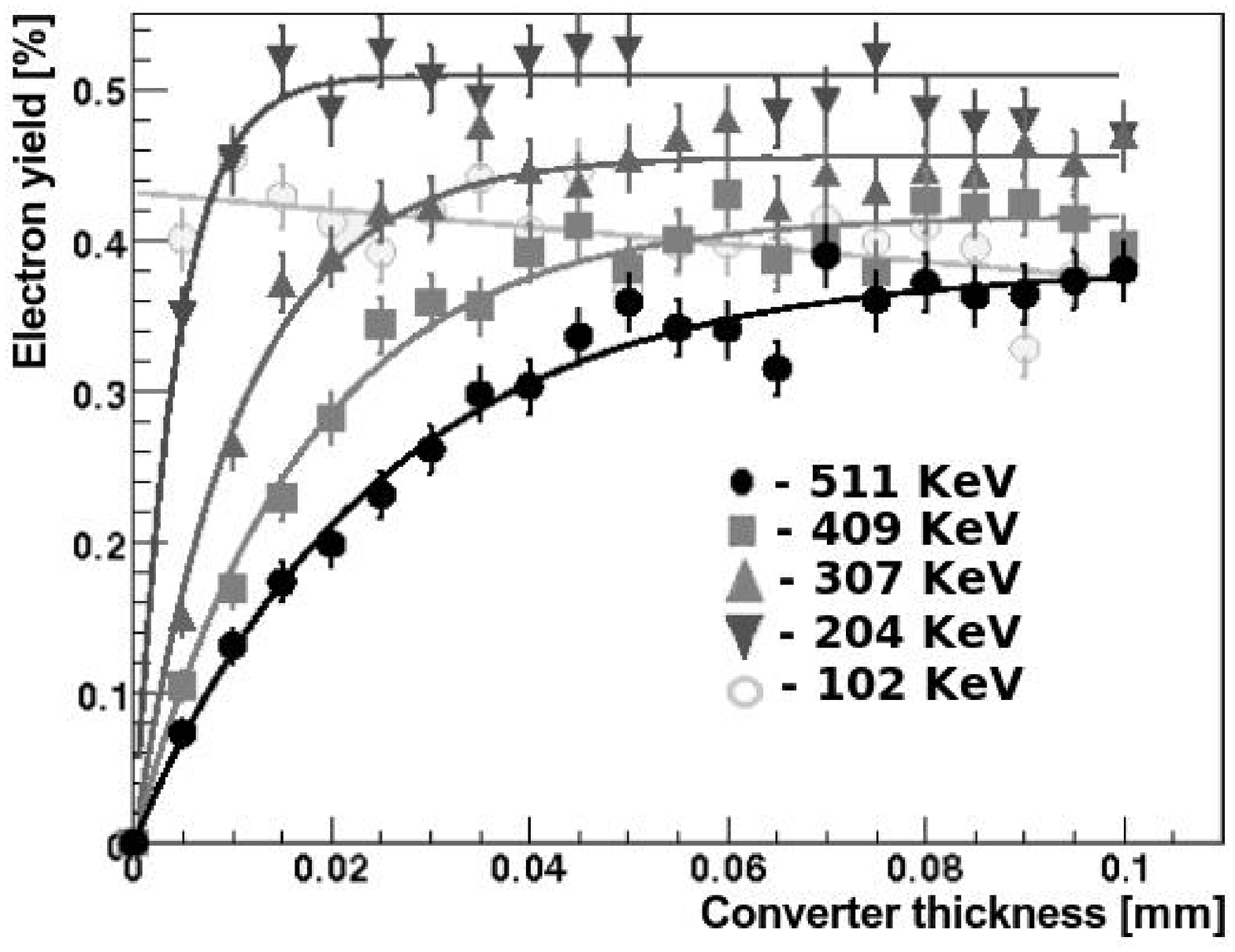}
    \caption{GC design -- electron yield  vs converter thickness: (a) for five converter materials and 511 $KeV$ photons;  (b)  for five input photon energies with Bi converter.}
  \label{GCyield}
  \end{center}
\end{figure}

\smallskip
The results for converter thicknesses in the range $(1\div100)\mu$m and five equidistant input photon energies are shown on Fig. \ref{GCyield}. The yield saturates at about $40 \mu m$, giving the maximum yield for each material. The maximum yield ranges $(0.30 \div 0.38)$\% and is highest for bismuth $(0.382 \pm 0.009)\%$, followed by lead $(0.380 \pm 0.007)\%$. The electron yield is indeed higher for the low energy (Compton-scattered) electrons than for 511 keV photons, in agreement with Eq. (\ref{thin_conv}). This result clearly disproves the intuitive gas-converter design.

\medskip\noindent
{\bf Gas-insulator-converter (GIC) design.} 
A bismuth converter sandwiched between two glass plates was used as a cathode in that RPC design (Fig. \ref{fig2}b). The clear design advantage is the absence of a direct contact between the gas and the converting material. The glass surface is highly resistive and smooth which prevents discharges. Furthermore, its aging is well studied \cite{glass}. The insulator between the converter and the gas effectively decreases the detector sensitivity towards scattered in the human body photons, as it absorbs them with higher probability, the drop becoming essential above  50 $\mu$m insulator thickness.

\begin{figure}[h!]
    \begin{center}
    \setlength{\unitlength}{1cm}   
    \includegraphics[height=5.5cm, width=6.5cm]{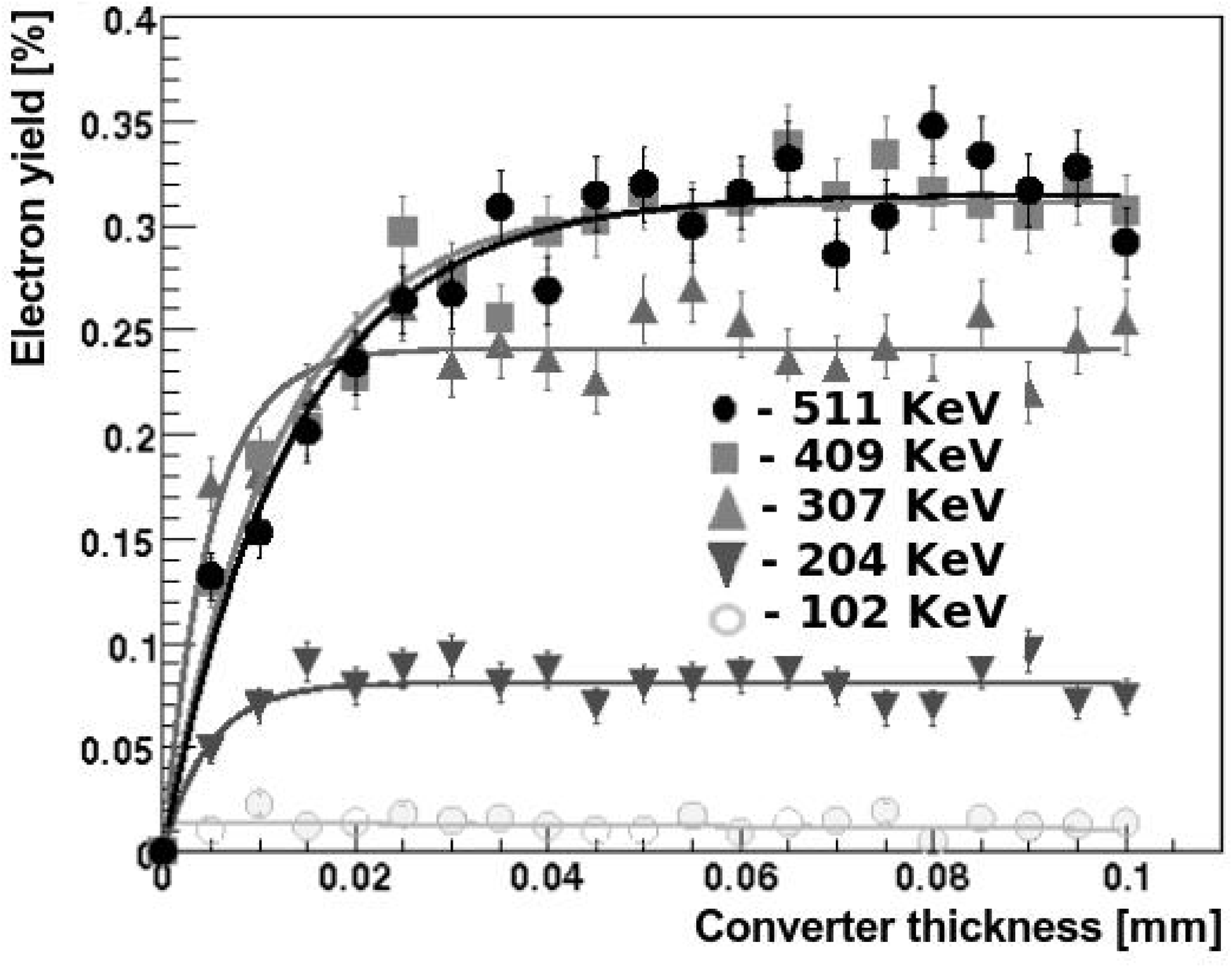}
    \includegraphics[height=5.5cm, width=6.5cm]{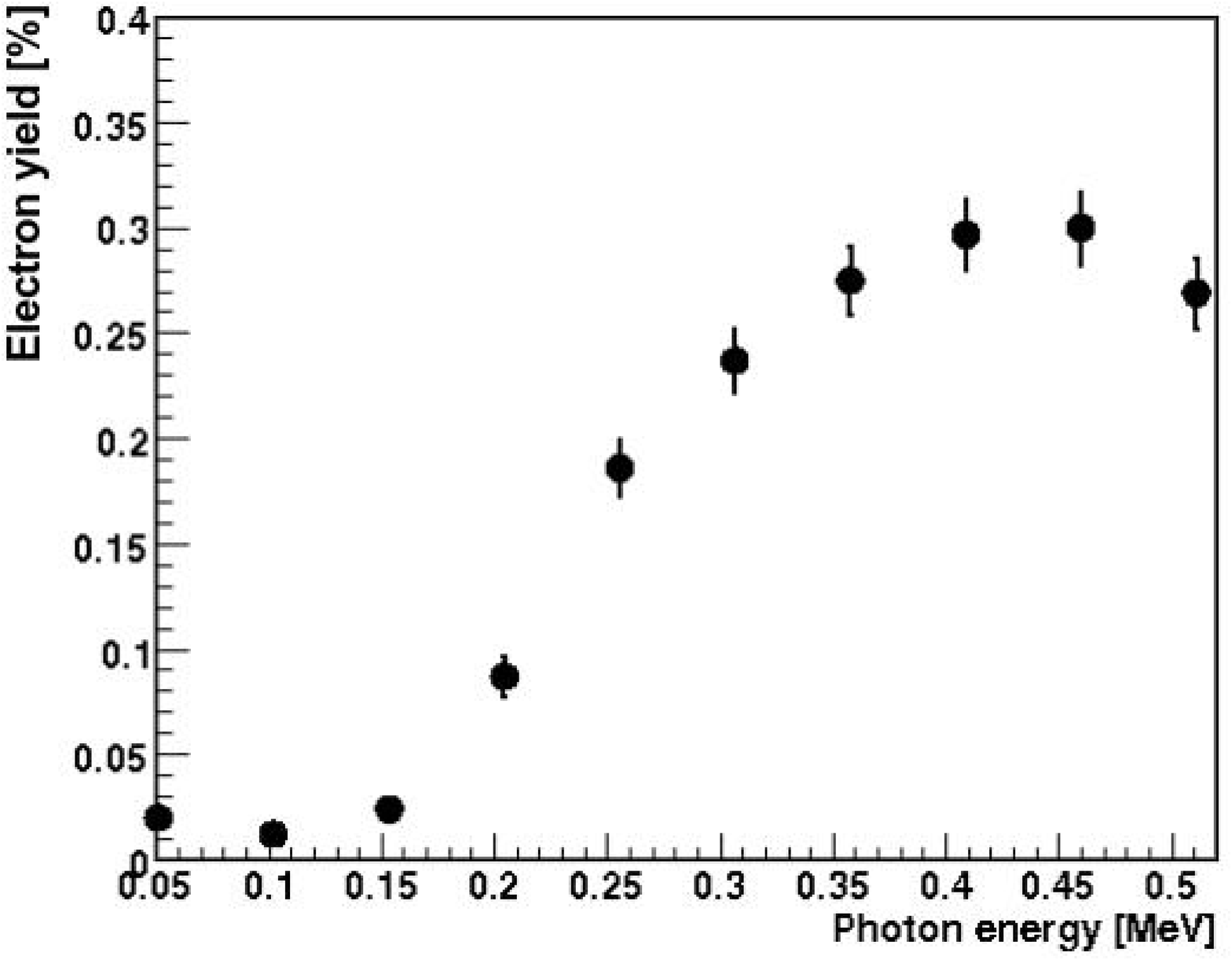}
    \caption{GIC design: (a) electron yield vs converter thickness for different photon energies, Bi converter; 
(b) electron yield as a function of the input photon energy (40 $\mu m$ Bi converter).}
  \label{GICyield}
  \end{center}
\end{figure}

\smallskip
For 200 $\mu$m glass insulator, the electron yield for 511 keV photons saturates at converter thickness of about 40 $\mu$m and is (0.228 $\pm$ 0.020)\%, Fig. \ref{GICyield}, which is 40\% lower than in the GC design. However, the electron yield for 307 keV photons  is  about 90\% lower than in the GC design --- (0.045 $\pm$ 0.008)\%, so indeed an effective suppression of the registration of Compton-scattered photons without energy discrimination is  achieved. The optimal parameters for the sandwich-type construction appear to be 50 $\mu$m thickness for both the Bi converter and the glass insulator.

\begin{figure}[ht]
\begin{center}
\includegraphics[width=.4\textwidth]{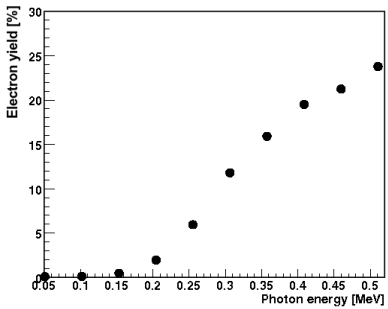}
\end{center}
\caption{Electron yield vs the input photon energy (detector response function; 100-gap MGIC design). Error bars are smaller than the data points.}
\label{fig3}
\end{figure}

\medskip\noindent
{\bf Multi-gap sandwich-type design.} 
A possibility to partly compensate the decrease in the electron yield due to the additional insulator included might be provided by a multi-gap detector design,  (Fig. \ref{fig2}c). The electron yield for each gap is given again by Eq. (\ref{thick_conv}), but with different initial numbers of photons for the different gaps, accounted in the coefficient $a_l$:
\beq\label{MGIC_coeff}
a_{l,i}=a_0e^{-(i-1)x/c},
\eeq
where $a_0$ is the $a_l$ parameter for the first gap, $x$ is the converter thickness for one gap, $c$ has the same meaning as in Eq. (\ref{thick_conv}). Thus, for $n$ gaps we get
\beq\label{N_gaps}
N=\sum_{i=1}^n a_0 \left(e^{-x/c}- e^{-x/b}\right) e^{-(i-1)x/c} = a_0  \left(e^{-x/c}- e^{-x/b}\right) \frac{1-e^{-(n-1)x/c}}{1-e^{-x/c}}.
\eeq

In fact, the enhancement of the detector efficiency is both because of the increased electron yield and due to the increased difference in the sensitivity for 511 keV photons and for lower-energy ones. We studied in detail a 100-gap stack, the entities being formed by glass (50 $\mu$m) -- Bi (50 $\mu$m) -- glass (50 $\mu$m) plates. The gas mixture was composed of 85\% C$_2$H$_2$F$_4$, 5\% i-C$_4$H$_{10}$, and 10\% SF$_6$. The avalanche development in the gas was not simulated in the study. The obtained electron yield of (23.8 $\pm$ 0.4)\% ensures photon efficiency comparable to that of the crystal PET detectors. The highest yield is observed for 511 keV photons, the sensitivity for photons with 307 keV is two times smaller. Electron yield vs the input photon energy is presented on Fig. \ref{fig3}.

\begin{figure}[t]
\begin{center}
\includegraphics[height=5.5cm, width=6.5cm]{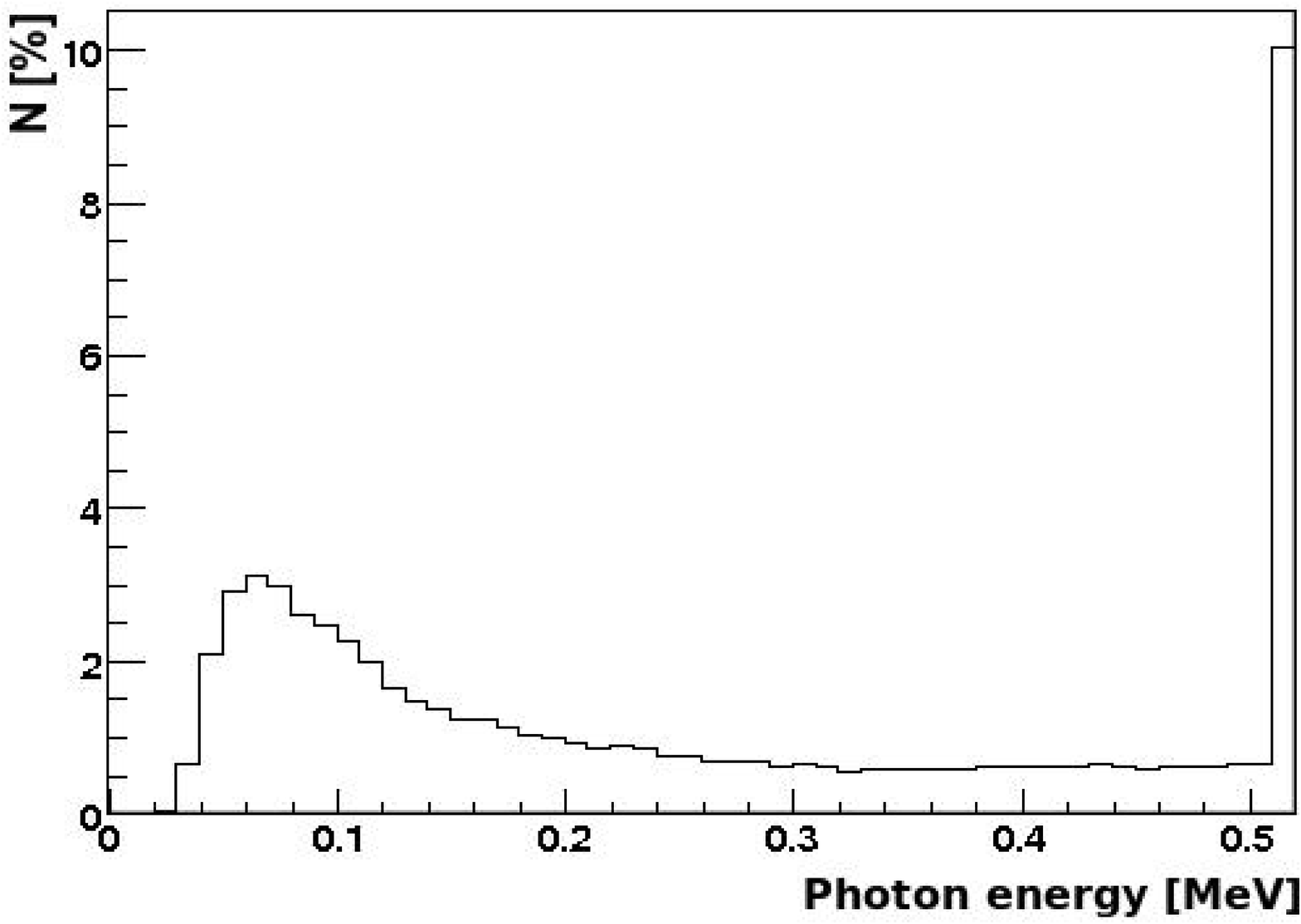}
\includegraphics[height=5.5cm, width=6.5cm]{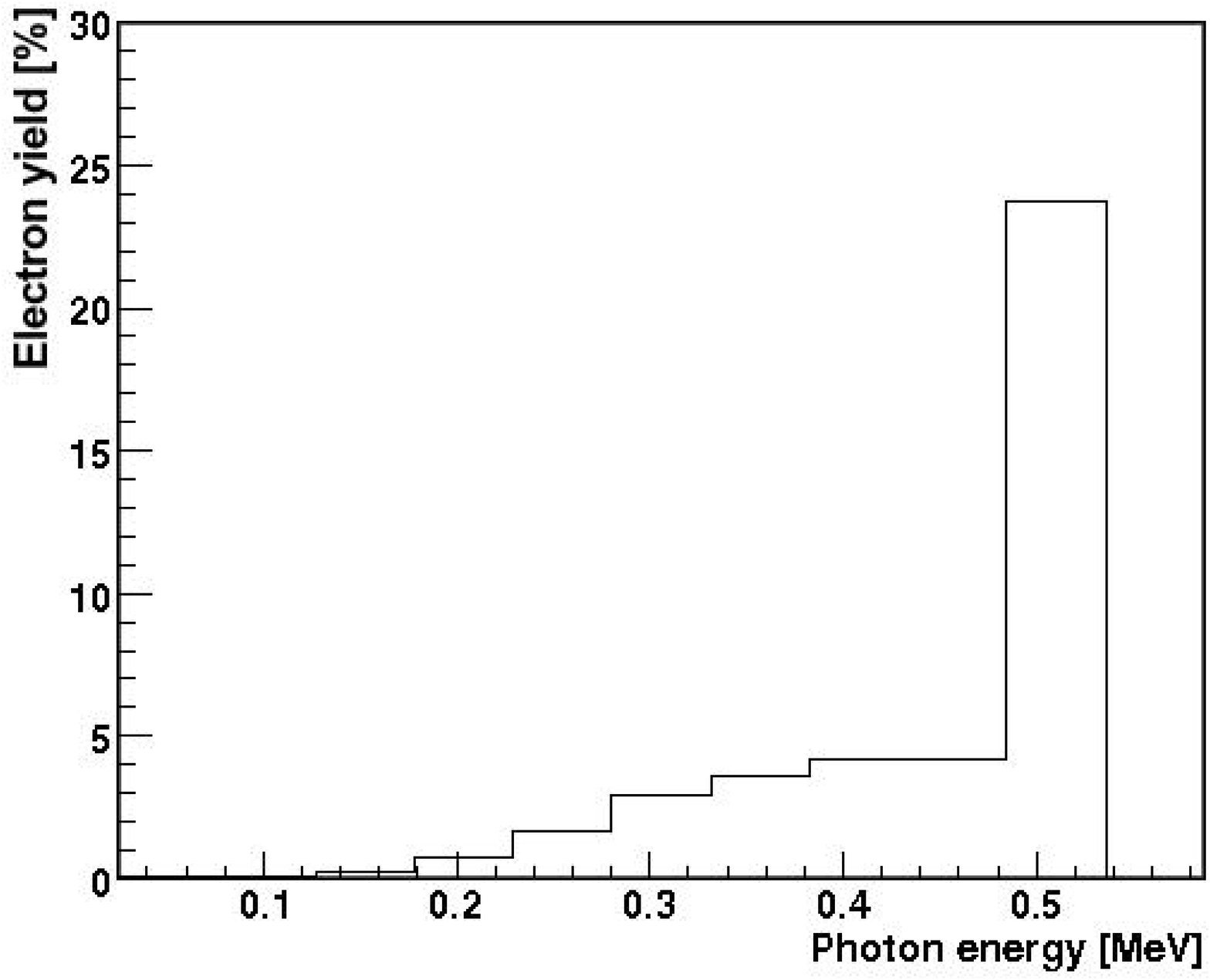}
\end{center}
\caption{(a) Energy spectrum ot the photons, leaving the human body; (b) Electron yield in the 100-gap MGIC detector for photons from a realistic human-body PET scan.}
\label{spectra}
\end{figure}

\smallskip
For simulation of the processes of interest in the human body, the latter was represented as a homogeneous parallelepiped of size $40\times40\times150 ~cm^3$, density of 1.01 $g/cm^3$, and the following contents: O -- 61.4\%;  C -- 22.9\%; H -- 10.0\%; N -- 2.6\%; Ca -- 1.4\%; P -- 1.1\%; K -- 0.2\%; S -- 0.2\%; Na -- 0.1\%; Cl -- 0.1\%. The photon propagation was considered as starting from the center of the volume. The energy spectrum ot the photons, leaving the human body, is presented on Fig. \ref{spectra}(a). This simulation does not account for the energy resolution curve of a particular detector (as e.g. in \cite{energy-res}) but only for the physical processes within the body. The convolution of this spectrum with the detector response is presented on Fig. \ref{spectra}(b): about  86\% of the registered in the PET process photons are with energies above 380 keV and are thus suitable for PET reconstruction.

\bigskip\noindent
{\bf RPC prototypes.}
Based on the simulation results we adopted the MGIC design in our prototypes. The basic module is a six-gap glass RPC with readout strips on both sides. Fishing lines are used for spacers. Graphite paint is applied to form the high voltage electrodes on the outermost glasses. The high voltage electrodes are insulated from the readout strips by a mylar foil. The copper readout strips are implemented on PCB boards. Twenty independent six-gap RPCs should be stacked one on top of the other to form 120 gas gaps supermodule.

\begin{figure}
\begin{center}
\includegraphics[trim= 0 0 0 -0.5cm, clip,width=.55\textwidth]{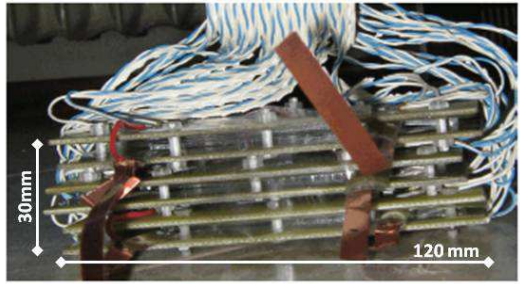}
\end{center}
\caption{The MGIC RPCPET prototype (small).}
\label{fig6}
\end{figure}

\smallskip
We constructed several small-size RPCPET prototypes and also a full-scale one, after having tried different techniques and materials, for example lead containing resistive paint, to form the converting layer between the two glasses. The final design of the small-size detector prototype is shown on Fig. \ref{fig6}.  The parameters of both small-size and full-scale prototypes are listed in the Table. Detailed analysis of the test results will be presented in a separate publication.

\begin{center}
\begin{tabular}{|l|c|c|}
\hline
& small-size prototype & full-scale prototype\\
\hline
number of gaps & 6 & 6\\
gas gap & 200  $\mu m$ & 200  $\mu m$\\
glass thickness & 150  $\mu m$ & 100  $\mu m$\\
dimensions & $ 120 mm \times 70 mm$ &   $ 350 mm \times 70 mm$\\
\hline
\end{tabular}
\label{tab:table1}
\end{center}

The intrinsic RPC noise is lower than 5 Hz/cm$^2$ and is negligible compared to the expected photon flux of about two thousand photons per cm$^2$. An important goal of our design is to achive time resolution of about 35 ps, which is essencial for TOF measurements. Our estimations show that detector timing resolution of 35 ps constrains the annihilation point to an 11 mm LOR region.


\section{Conclusions}

We performed model investigations towards the design of an RPC-based PET detector encompassing the whole chain from the annihilation of the positrons in the body, through the conversion of the created photons into electrons and to the optimization of the electron yield in the gas. Contrary to the intuitive expectations, a direct contact between the converter and the gas does not provide sufficient Compton suppression. As a result, a multi-gap sandwich-type gas-insulator-converter detector with Bi or Pb as converter materials was chosen. In this case, the detector efficiency for registration of 511 keV photons was shown to reach values at the level of  24\%, in parallel with a significantly suppressed response to Compton-scattered lower-energy photons. This new feature is crucial for the construction of RPC-based PET detectors. In particular, a device comprising 100 individual RPC gaps will ensure signal-to-background ratio (511 keV photons to photons with energies below 380 keV) better than 6:1.
 
\section{Acknowledgments}
This work was supported by Bulgarian Science Fund under contract number  DO 02-183/2008. We acknowledge the useful comments and discussions with Crispin Williams, Paolo Vitulo, Sergio Ratti and Marec Palka.


\begin{thebibliography}{9}

\bibitem{pet} Michael E. Phelps, \emph{PET Molecular Imaging and Its Biological Applications} (2004, Springer--Verlag New York; ISBN 0--387--40359--0)

\bibitem{sipm} D. Henseler et al, \emph{SiPM performance in PET applications: An experimental and theoretical analysis}, \href{http://dx.doi.org/10.1109/NSSMIC.2009.5402157}{\emph{Nuclear Science Symposium Conference Record (NSS/MIC), 2009 IEEE } (2009) 1095--1948}

\bibitem{axpet} E. Bolle et al, \emph{A demonstrator for a new axial PET concept}, \href{http://dx.doi.org/10.1109/NSSMIC.2008.4774326}{\emph{Nuclear Science Symposium Conference Record, 2008. NSS '08. IEEE} (2008) 4571--4574} 

\bibitem{petctpetmri} B. Pichler et al, \emph{Multimodal imaging approaches: PET/CT and PET/MRI}, \href{http://dx.doi.org/10.1007/978-3-540-72718-7_6}{\emph{Handb Exp Pharmacol.} {\bf 185} (2008) 109--32}

\bibitem{petmri} N.E. Bolus et al, \emph{PET/MRI: The Blended-Modality Choice of the Future?}, \href{http://dx.doi.org/10.2967/jnmt.108.060848}{\emph{J. Nucl. Med. Technol.} {\bf 37} (2009) 263--271} 

\bibitem{dSiPM} C. Degenhardt et al., \emph{The digital Silicon Photomultiplier -- A novel sensor for the detection of scintillation light }, \href{http://dx.doi.org/10.1109/NSSMIC.2009.5402190}{\emph{Nuclear Science Symposium Conference Record (NSS/MIC), 2009 IEEE} (2009) 2383--2386} 

\bibitem{santonico}  R. Santonico and R. Cardarelli, \emph{Development of Resistive Plate Counters}, \href{http://dx.doi.org/10.1016/0029-554X(81)90363-3}{\emph{Nucl. Instr. Meth.}, {\bf 187}, (1981) 377--380}



\bibitem{fonte2}  P. Fonte, \emph{Applications and New Developments in Resistive Plate Chambers}, \href{http://dx.doi.org/10.1109/TNS.2002.1039583}{\emph{IEEE Transactions on Nuclear Science}, {\bf 49} (2002) 881--887} 

\bibitem{rpcpet1} A. Bianco et al, \emph{RPC-PET: A New Very High Resolution PET Technology}, \href{http://dx.doi.org/10.1109/TNS.2006.876005}{\emph{IEEE Transactions on Nuclear Science}, {\bf 53}, (2006) 2489--2494} 

\bibitem{rpcpet2} G. Belli et al, \emph{RPC: from High Energy Physics to Positron Emission Tomography}, \href{http://dx.doi.org/10.1088/1742-6596/41/1/068}{\emph{J. Phys.: Conf. Ser.}{\bf 41} (2006) 555}

\bibitem{crispin} C. Williams et al, \emph{A 20 ps timing device-A Multigap Resistive Plate Chamber with 24 gas gaps}, \href{http://dx.doi.org/10.1016/j.nima.2008.06.013}{\emph{Nucl. Instr. Meth. A}  {\bf 594} (2008) 39--43}

\bibitem{cms} M. Abbrescia et al, \emph{Cosmic ray tests of double-gap resistive plate chambers for the CMS experiment}, \href{http://dx.doi.org/10.1016/j.nima.2005.06.074}{\emph{Nucl. Instrum. Meth. A} {\bf 550} (2005) 116--126}

\bibitem{atlas} G. Chiodini et al, \emph{RPC cosmic ray tests in the ATLAS experiment}, \href{http://dx.doi.org/10.1016/j.nima.2007.07.080}{\emph{Nucl. Instrum. Meth. A} {\bf 581} (2007) 213--216}

\bibitem{fonte1} A.Blanco et al., \emph{Efficiency of RPC detectors for whole-body human TOF-PET}, \href{http://dx.doi.org/10.1016/j.nima.2008.12.134} {\emph{Nucl. Instr. Meth. A}, {\bf 602}, (2009) 780--783}

\bibitem{geant4_1} S. Agostinelli et al, \emph{Geant4-a simulation toolkit} \href{http://dx.doi.org/10.1016/S0168-9002(03)01368-8}{\emph{Nucl. Instrum. Meth. A} {\bf 506} (2003) 250--303} 

\bibitem{geant4_2} K. Amako et al, \emph{Geant4 developments and applications} \href{http://dx.doi.org/10.1109/TNS.2006.869826}{\emph{IEEE Transactions on Nuclear Science} {\bf 53 No. 1} (2006) 270--278}

\bibitem{hppc} C. D'Ambrosio et al, \emph{A Hybrid Parallel Plate gas Counter for medical imaging}, \href{http://dx.doi.org/10.1016/j.nima.2006.10.193}{\emph{Nucl. Instrum. Meth. A} {\bf 572} (2007) 244--245}

\bibitem{glass} C. Gustavino et al, \emph{Ageing and recovering of glass RPC}, \href{http://dx.doi.org/10.1016/j.nima.2004.07.012}{\emph{Nucl. Instrum. Meth. A} {\bf 533} (2004) 116--120} \href{http://dx.doi.org/10.1063/1.3322562}{\emph{AIP Conf. Proc.} {\bf 1203} (2009) 820--825}

\bibitem{energy-res} L.M. Popescu et al, \emph{PET energy-based scatter estimation and image reconstruction with energy-dependent corrections}, \href{dx.doi.org/10.1088/0031-9155/51/11/016}{\emph{Phys. Med. Biol.}{\bf 51} (2006) 2919--2937}

\end{thebibliography}
\end{document}